\documentclass[11pt]{article}
\usepackage{amsmath}
\usepackage{graphicx}
\usepackage{color}
\usepackage{orcidlink}
\usepackage{hyperref}
\hypersetup{colorlinks=true, linkcolor=blue, citecolor=blue, urlcolor=blue}
\begin{document}
\baselineskip=14 pt

\begin{center}
    {\large {\bf Cosmological constant Petrov type-N space-time in Ricci-inverse gravity }}
\end{center}

\vspace{0.3cm}

\begin{center}
     {\bf F. Ahmed\orcidlink{0000-0003-2196-9622}}\footnote{\bf faizuddinahmed15@gmail.com}\\
    \vspace{0.1cm}
    {\it Department of Physics, University of Science \& Technology Meghalaya,Ri-Bhoi, 793101, India}\\
    \vspace{0.5cm}
    {\bf J. C. R. de Souza\orcidlink{0000-0002-7684-9540}}\footnote{\bf jean.carlos@fisica.ufmt.br} and {\bf A. F. Santos\orcidlink{0000-0002-2505-5273}}\footnote{\bf alesandroferreira@fisica.ufmt.br (Corresp. author)}\\
    \vspace{0.1cm}
    {\it Instituto de Física, Universidade Federal de Mato Grosso, Cuiabá, Mato Grosso 78060-900, Brazil}
\end{center}

\vspace{0.5cm}

\begin{abstract}
Our focus is on a specific type-N space-time that exhibits closed time-like curves in general relativity theory within the framework of Ricci-inverse gravity model. The matter-energy content is solely composed of a pure radiation field, and it adheres to the energy conditions while featuring a negative cosmological constant. One of the key findings in this investigation is the non-zero determinant of the Ricci tensor ($R_{\mu\nu}$), which implies the existence of an anti-curvature tensor ($A^{\mu\nu}$) and, as a consequence, an anti-curvature scalar ($A \neq R^{-1}$). Furthermore, we establish that this type-N space-time serves as a solution within modified gravity theories via the Ricci-inverse model, which involves adjustments to the cosmological constant ($\Lambda$) and the energy density ($\rho$) of the radiation field expressed in terms of a coupling constant. As a result, our findings suggest that causality violations remain possible within the framework of this Ricci-inverse gravity model, alongside the predictions of general relativity.  
\end{abstract}

\vspace{0.2cm}


\vspace{0.2cm}
{\bf PACS number(s):} 04.20.Jb; 04.50.Kd; 98.80.Es;

\section{Introduction}

 The solutions of Einstein's field equations that exhibit intriguing features have garnered acceptance within the scientific community. These solutions are systematically categorized using the Petrov classification scheme. Since curvature is a localized property of space-time, the Petrov type provides insight into the local algebraic characteristics of the space-time geometry. Following this scheme \cite{AZP,AZP2} involves first establishing a set of null tetrad vector fields denoted as $(\mathbf{k}, \mathbf{l}, \mathbf{m}, \bar{\mathbf{m}})$ for a given space-time and subsequently determining the Weyl scalars $\Psi_{i}$ ($i=0,1,2,3,4$) \cite{HS,JB}. In the context of Petrov type-N space-time, the only non-zero Weyl scalar is $\Psi_4 \neq 0$, and the Weyl tensor satisfies the condition $C_{\mu\nu\rho\sigma}\,k^{\sigma}=0$, where $k^{\sigma}$ is the quadruple principal null direction (PND) along which the gravitational radiation propagates. The Petrov classification of the Weyl scalars and the asymptotic forms of radiative fields originating from spatially bounded sources underscores the fundamental role played by type-N solutions in gravitational radiation theory (for a comprehensive review, see \cite{JB2,JB3}). All solutions of the vacuum Einstein equations, including those with a non-zero cosmological constant $\Lambda$, falling under type N and featuring a non-twisting null geodesic congruence, are well-established \cite{HS,JB,AGD,JB4,JB5,JB6,JP,SBE,DS1,AOP}. Additional non-vacuum solutions, both without and with a cosmological constant, within the framework of Petrov type-N space-times with non-twisting, non-expanding, and shear-free geodesic null congruence, are constructed in \cite{DS2,PTEP,refa10}. Recent investigations have extended to include twisting type-N vacuum solutions with a nonzero cosmological constant \cite{XZ,XZ2}. Furthermore, type N universal space-time has also attracted attention in recent times (see, refs. \cite{SH,SH2,SH3}). In type-N space-time without twisting, a non-expanding and shear-free geodesic null vector field emerges as the conduit for gravitational wave propagation. These gravitational wave space-times are either plane-fronted gravitational waves with parallel rays (called pp-wave for non-vacuum solutions of the field equations) or plane-wave (vacuum solutions) space-times, belonging to the Kundt class. The confirmation of gravitational waves, or ripples in space and time, resulting from the merger of black holes by the LIGO scientific collaboration in 2015, validated the success of general relativity in contemporary physics. Thus, Petrov type-N space-time holds significant importance in the realm of gravitational wave theory.

One of the biggest problems in modern cosmology is the current accelerated expansion of the universe. There is strong observational data confirming this phenomenon \cite{Riess, Perm, WMAP}. One possibility to understand this accelerated phase is add a new exotic component, called dark energy. This component has a negative pressure that causes gravity to behave repulsively on large cosmological scales \cite{DE}. The best-known candidate for dark energy is the cosmological constant $\Lambda$. Taking this component and assuming the existence of dark matter, the $\Lambda$CDM model has been proposed. Although it is a very popular model and can successfully explain the observational results, it is not without problems. The main problem is the famous cosmological constant problem \cite{Wein}. Another way to explain the cosmological observations is to propose modifications to Einstein's general relativity \cite{Mod1, Mod2, Mod3, Mod4, Mod5}. The simplest manner to construct alternative theories of general relativity is to include an additional term in the Einstein-Hilbert Lagrangian or to modify the structure of the Lagrangian which implies modifying the Ricci scalar itself. Among the various modified gravity theories existing in the literature, there is a recent model, called Ricci-inverse gravity, which has attracted attention. In this paper, this model is considered.

Ricci-inverse gravity \cite{RIG} is an alternative theory to general relativity that modifies the Einstein-Hilbert action by introducing a geometrical object called anti-curvature scalar $A$. The anti-curvature scalar is the trace of the anti-curvature tensor denoted as $A^{\mu\nu}$, which is defined as the inverse of the Ricci tensor $R^{\mu\nu}$. The anti-curvature tensor satisfies the condition $A^{\mu\sigma}R_{\sigma\nu}=\delta^\mu_\nu$. It is important to note that the anti-curvature scalar $A$ is not the inverse of the Ricci scalar $R$. There are two classes that define this model: class I, characterized by the function $f(R, A)$ that depends on Ricci and anti-curvature scalars, and class II, given by the function $f(R, A^2)$ which depends on the square of the anti-curvature tensor. Here it is considered class I. In recent years, some investigations have been developed using this gravitational theory. For example,  no-go theorem for inflation has been investigated \cite{Do, Do1}, cosmic structure has been studied \cite{Cosmic},  evolution from matter-dominated epoch to accelerated expansion epoch has been analyzed \cite{Dasa}, the matter-antimatter asymmetry through baryogenesis in the realm of $f(R, A)$ theory of gravity has been discussed \cite{Bar}, the causality issue using an axially symmetric space-time has been investigated \cite{Our}, anisotropic stellar structures has been explored \cite{refa6}, among others applications.

In this paper, we aim to investigate a type-N space-time in general relativity,  possessing a non-twisting, non-expanding, and shear-free geodesic null congruence within the framework of Ricci-inverse gravity. For that we consider a specific example: pure radiation field space-time characterized by a negative cosmological constant that was investigated in \cite{refa10}.  This space-time represents pp-wave non-vacuum solution of the field equations satisfying the energy conditions. In general relativity, this solution represents an exact solution of the field equations and exhibits closed time-like curves at an instant of time, thus violating the causality condition.  Our study reveals that this also serve as a solution within the framework of Ricci-inverse gravity, where the cosmological constant $\Lambda$ is replaced by a modified value $\Lambda_{\mbox{mod}}$, and the energy-density of the radiation field, $\rho$, is altered to $\rho_{\mbox{mod}}$, which is connected to a coupling parameter. Consequently, we demonstrate that the violation of causality is permitted in this modified gravity theory as well. It is worthwhile mentioning that this particular type-N space-time is a solution in Ricci-inverse gravity since the determinant of the Ricci tensor is nonzero, $\det(R_{\mu\nu})\neq 0$ which ensures the existence of an anti-curvature tensor $A^{\mu\nu}$ which is symmetric in nature analogue to the metric tensor or the Ricci tensor.

This paper is organized as following. In Section 2, the cosmological constant type-N space-time with causality violation is introduced. Considering pure radiation as matter content,  it is shown that this metric is the solution in general relativity obtaining a negative cosmological constant and a positive energy density. Next, this metric is investigated in Ricci-inverse gravity. Some conclusions and remarks are presented in Section 3.

\section{Cosmological constant type-N space-time with causality violation in Ricci-inverse gravity}

In this section, we examine an example of a type-N pure radiation field solution derived from Einstein's field equations in the context of general relativity. The matter content in this scenario consists of a pure radiation field that satisfies the energy conditions and includes a negative cosmological constant. Notably, this type-N space-time permits the existence of closed time-like curves within time-like regions, thus violating the causality condition. We take this type-N space-time and investigate it within the framework of modified gravity theories using the Ricci-inverse approach.

The line-element that describes this type-N space-time in cylindrical coordinates ($t, r, \phi, z$) is given by the following expression (assuming $c=1$ and $8\pi G=1$) \cite{refa10}:
\begin{equation}
    ds^2=g_{rr}\,dr^2+2\,g_{t\phi}\,dt\,d\phi+g_{\phi\phi}\,d\phi^2+2\,g_{z\phi}\,dz\,d\phi+g_{zz}\,dz^2,\label{1}
\end{equation}
with different components of the metric tensor $g_{\mu\nu}$ given by 
\begin{eqnarray}
    &&g_{t\phi}=-\frac{1}{2}\,\cosh t\,\sinh^2 (\alpha\,r),\nonumber\\
    &&g_{rr}=\mbox{coth}^2 (\alpha\,r),\nonumber\\
    &&g_{\phi\phi}=-\sinh t\,\sinh^2 (\alpha\,r),\nonumber\\
    &&g_{zz}=\sinh^2 (\alpha\,r),\nonumber\\
    &&g_{z\phi}=\beta\,z\,\sinh^2 (\alpha\,r),\label{2}
\end{eqnarray}
where $\alpha>0, \beta>0$ are arbitrary positive constants.

To convert this metric (\ref{1}) with (\ref{2}) into a standard form, we perform transformations as follows 
 \begin{equation}
    t \to \sinh^{-1} (\tau),\quad r \to \frac{1}{\alpha}\,\sinh^{-1} (\alpha\,\varrho),\label{3}
\end{equation}
into the metric (\ref{1}), we obtain the following line-element in the chart $(\tau, \varrho, \phi, z)$ given by 
\begin{equation}
    ds^2=\frac{d\varrho^2}{\alpha^2\,\varrho^2}+\alpha^2\,\varrho^2\,\Big(-d\tau\,d\phi-\tau\,d\phi^2+2\,\beta\,z\,dz\,d\phi+dz^2\Big).\label{4}
\end{equation}

Our primary objective is to investigate this metric (\ref{4}) within the context of Ricci-inverse gravity. This choice is motivated by the fact that the determinant of the Ricci tensor for the space-time described by the metric (\ref{4}) is non-zero, as we will discuss in detail below. Before that, we first review this metric in the context of general relativity and then will investigate within the frame of modified theories of gravity. 

The covariant metric tensor $g_{\mu\nu}$ and its contravariant form $g^{\mu\nu}$ for the metric (\ref{4}) are given by
\begin{eqnarray}
     &&g_{\mu\nu}=\begin{pmatrix}
         0 & 0 & -\frac{\alpha^2\,\varrho^2}{2} & 0\\
         0 & \frac{1}{\alpha^2\,\varrho^2} & 0 & 0\\
         -\frac{\alpha^2\,\varrho^2}{2} & 0 & -\alpha^2\,\varrho^2\,\tau & \beta\,z\,\alpha^2\,\varrho^2\\
         0 &  0 & \beta\,z\,\alpha^2\,\varrho^2 & \alpha^2\,\varrho^2
     \end{pmatrix},\nonumber\\ 
     &&g^{\mu\nu}=\begin{pmatrix}
         \frac{4\,(\tau+\beta^2\,z^2)}{\alpha^2\,\varrho^2} & 0 & -\frac{2}{\alpha^2\,\varrho^2} & \frac{2\,\beta\,z}{\alpha^2\,\varrho^2}\\
         0 & \alpha^2\,\varrho^2 & 0 & 0\\
         -\frac{2}{\alpha^2\,\varrho^2} & 0 & 0 & 0\\
         \frac{2\,\beta\,z}{\alpha^2\,\varrho^2} &  0 & 0 & \frac{1}{\alpha^2\,\varrho^2}
     \end{pmatrix}. \label{5}
\end{eqnarray}
The covariant Ricci tensor $R_{\mu\nu}$ and its contravariant form for the metric (\ref{4}) are given  
\begin{eqnarray}
    &&R_{\mu\nu}=\begin{pmatrix}
        0 & 0 & \frac{3\,\alpha^4\,\varrho^2}{2} & 0\\
        0 & -\frac{3}{\varrho^2} & 0 & 0\\
        \frac{3\,\alpha^4\,\varrho^2}{2} & 0 & \beta+3\,\tau\,\alpha^4\,\varrho^2 & -3\,\alpha^4\,\beta\,z\,\varrho^2\\
        0 & 0 & -3\,\alpha^4\,\beta\,z\,\varrho^2 & -3\,\alpha^4\,\varrho^2
    \end{pmatrix},\nonumber\\
    &&R^{\mu\nu}=\begin{pmatrix}
        \frac{4}{\varrho^4}\,\big(\frac{\beta}{\alpha^4}-3\,\varrho^2\,(\tau+\beta^2\,z^2)\big) & 0 & \frac{6}{\varrho^2} & -\frac{6\,\beta\,z}{\varrho^2}\\
        0 & -3\,\alpha^4\,\varrho^2 & 0 & 0\\
        \frac{6}{\varrho^2} & 0 & 0 & 0\\
        -\frac{6\,\beta\,z}{\varrho^2} & 0 & 0 & -\frac{3}{\varrho^2}
    \end{pmatrix}. \label{6}
\end{eqnarray}
Finally, the Ricci scalar for metric (\ref{4}) is given by
\begin{equation}
    R=g_{\mu\nu}\,R^{\mu\nu}=-12\,\alpha^2.\label{7}
\end{equation}

The non-zero components of the Einstein tensor $G_{\mu\nu}$ are
\begin{eqnarray}
    &&G_{\tau\phi}=-\frac{3\,\alpha^4\,\varrho^2}{2}=3\,\alpha^2\,g_{\tau\phi},\nonumber\\
    &&G_{\varrho\varrho}=\frac{3}{\varrho^2}=3\,\alpha^2\,g_{\varrho\varrho},\nonumber\\
    &&G_{\phi\phi}=-3\,\tau\,\alpha^4\,\varrho^2+\beta=3\,\alpha^2\,g_{\phi\phi}+\beta,\nonumber\\
    &&G_{\phi z}=3\,\alpha^4\,\varrho^2\,\beta\,z=3\,\alpha^2\,g_{\phi z},\nonumber\\
    &&G_{zz}=3\,\alpha^4\,\varrho^2=3\,\alpha^2\,g_{zz}\,.\label{a}
\end{eqnarray}

The Einstein field equations with a cosmological constant and the energy-momentum tensor $\mathcal{T}^{\mu\nu}$ are given by
\begin{equation}
    G_{\mu\nu}+\Lambda\,g_{\mu\nu}=\mathcal{T}_{\mu\nu},\label{8}
\end{equation}
where the right hand side is given by
\begin{equation}
    \mathcal{T}_{\mu\nu}=\rho\,k_{\mu}\,k_{\nu},\quad \mathcal{T}^{\mu}_{\,\,\mu}=0,\label{8a}
\end{equation}
with $\rho$ being the energy-density of a pure radiation field, and $k_{\mu}=\delta^{\phi}_{\mu}=(0,0,1,0)$ is a null vector that satisfies the relation $k^{\mu}\,k_{\mu}=0$. The non-zero component of the energy-momentum tensor is $\mathcal{T}_{\phi\phi}=\rho$.

Using the metric tensor $g_{\mu\nu}$ given by (\ref{5}), the Einstein tensor $G_{\mu\nu}$ by (\ref{a}), and the energy-momentum tensor (\ref{8a}) into the equations (\ref{8}), one will find the following physical quantities given by
\begin{eqnarray}
    &&\Lambda=-3\,\alpha^2,\nonumber\\
    &&\rho=\beta>0.\label{10}
\end{eqnarray}

Thus, space-time (\ref{4}) is an exact solution of the field equations in general relativity with matter content a pure radiation field having constant energy-density and a negative cosmological constant.  To determine type of the chosen space-time (\ref{4}) using the Petrov classification scheme, one can use the Newman–Penrose formalism and construct a set of null tetrad vectors $(\bf k, l, m, \bar{m})$ \cite{HS}. One can easily show that the only Weyl scalars $\Psi_4 \neq 0$ and the rest are all equal to zero, which indicates that the chosen metric is of the Petrov type-N. The null vector field ${\bf k}$ satisfies the geodesic null congruence, that is, $k_{\mu;\nu}\,k^{\nu}=0$. Thus, the space-time we have chosen here admits a non-expanding, non-twisting, and shear-free geodesic congruence.  

Now, we show that this space-time admits closed causal curves. For that, let us consider curves defined by
\begin{equation}
    \varrho=\varrho_0,\quad z=z_0,\label{a1}
\end{equation}
where $\varrho_0, z_0$ are constants. Therefore, the space-time (\ref{4}) reduces to 2D Misner-like space \cite{Ori,Ori2} given by
\begin{equation}
    ds^2_{conf-Misner}=\Omega_0\,(-d\tau\,d\phi-\tau\,d\phi^2),\label{a2}
\end{equation}
where $\Omega_0=\alpha^2\,\varrho^2_{0}$ is the conformal constant factor. 

The Misner space is a $2D$ space-time with the metric \cite{Mis}
\begin{equation}
    ds^2_{Misner}=-2\,dT\,d\psi-T\,d\psi^2,\label{a3}
\end{equation}
where $-\infty <T < +\infty$ and the coordinate $\psi$ is periodic, that is, $\psi \to \psi +2\,n\,\pi$ with $n=0,\pm\,1,\pm\,2,....$. The curves $T=const=T_0$ are all closed due to the periodicity of $\psi$. The curves $T_0<0$ are spacelike and $T_0>0$ are time-like. It then follows that all points at $T_0>0$ rest on closed time-like curves (CTCs) but those at $T_0<0$ do not. Hence, the Misner space admits closed time-like at an instant of time, $T=const=T_0>0$. 

In our case, for the chosen space-time (\ref{4}), the closed curves defined by $\{\varrho, \phi, z\} \to \{\varrho_0, \phi+2\,n\,\pi, z_0\}$ is spacelike for $\tau<0$ and time-like for $\tau>0$. Hence, our four-dimensional space-time (\ref{4}) admits closed time-like curves at an instant of time analogue to the two-dimensional Misner space. These closed time-like curves evolve from an initial spacelike hypersurface in a causally well-behaved manner \cite{Ori2}. This point is clear from the following discussion. The metric component $g^{\tau\tau}$ for the metric (\ref{4}) is given by
\begin{equation}
    g^{\tau\tau}=\frac{4\,(\tau+\beta^2\,z^2)}{\alpha^2\,\varrho^2}\,.\label{a4}
\end{equation}
In the constant $z$-plane defined by $z=z_0=0$, we obtain
\begin{equation}
    g^{\tau\tau}=\frac{4\,\tau}{\alpha^2\,\varrho^2}\,.\label{a5}
\end{equation}
Thus, the hypersurface $\tau=const$ is spacelike for $\tau>0$ since $g^{\tau\tau}>0$ and time-like for $\tau<0$. The curve $\tau=0$ is null and serve as chronology horizon, that is, the hypersurface separating the causal and non-causal parts of space-time.

Thus, the Petrov type-N space-time (\ref{4}) which is a non-vacuum solution of the field equations admits closed time-like curves analogue to the Misner space. Now, we study this line-element (\ref{4}) in the context of modified theories of gravity via the Ricci-inverse gravity by introducing an anti-curvature tensor into the Lagrangian of the system, as discussed in Refs. \cite{RIG,Dasa}.

The action that describes the Ricci-inverse gravity is given as
\begin{eqnarray}
    S= \int d^4x \sqrt{-g}\left[(R + \kappa\,A-2\,\Lambda)+{\cal L}_m\right],\label{11}
\end{eqnarray}
where $g$ is the metric determinant, $\kappa$ is the coupling constant, $R=g_{\mu\nu}\,R^{\mu\nu}$ is the Ricci scalar, $A=g_{\mu\nu}\,A^{\mu\nu}$ is anti-curvature scalar, $\Lambda$ is the cosmological constant, and ${\cal L}_m$ is the matter Lagrangian. Varying the action (\ref{11}) with respect to the metric, the field equations that describe this gravitational theory is \cite{Our}
\begin{eqnarray}
   R^{\mu \nu} - \frac{1}{2}\,R\,g^{\mu \nu}+ \Lambda\,g^{\mu \nu} + M^{\mu \nu} = \mathcal{T}^{\mu \nu}.
\end{eqnarray}
Using the value of the Ricci scalar given in (\ref{7}), the last equation becomes
\begin{eqnarray}
R^{\mu \nu} +(6\,\alpha^2+\Lambda)\,g^{\mu \nu} + M^{\mu \nu} = \mathcal{T}^{\mu \nu},\label{12}
\end{eqnarray}
with $\mathcal{T}^{\mu\nu}$ being the standard energy-momentum tensor and the tensor $M^{\mu \nu}$ is defined as
\begin{eqnarray}
     M^{\mu\nu}&=&-\kappa\,\Big(A^{\mu\nu}+\frac{A}{2}\,g^{\mu\nu}\Big)\nonumber\\
     &+&\frac{\kappa}{2}\,\Big\{2\,g^{\kappa\mu}\nabla_{\iota}\nabla_{\kappa} (A^{\iota}_{\sigma}\,A^{\nu\sigma})-\nabla^2(A^{\mu}_{\iota}\,A^{\nu\iota})-g^{\mu\nu}\nabla_{\kappa}\nabla_{\iota}(A^{\kappa}_{\sigma}\,A^{\iota\sigma})\Big)\Big\}.\label{13}
\end{eqnarray}

The determinant of the metric tensor $g_{\mu\nu}$ and the Ricci tensor $R_{\mu\nu}$ for the space-time (\ref{4}) are given by
\begin{eqnarray}
    \mbox{det}\,g_{\mu\nu}=-\frac{\alpha^4\,\varrho^4}{4},\quad \mbox{det}\,R_{\mu\nu}=-\frac{81\,\alpha^{12}\,\varrho^4}{4}.\label{14}
\end{eqnarray}
Since the determinant of the Ricci tensor is non-zero, there exist an anti-curvature tensor $A^{\mu\nu}$ defined by 
\begin{eqnarray}
    A^{\mu\nu}=R^{-1}_{\mu\nu}=\frac{\mbox{adj}\,(R_{\mu\nu})}{\mbox{det}\,(R_{\mu\nu})}.\label{15}
\end{eqnarray}

For the space-time (\ref{4}), this anti-curvature tensor $A^{\mu\nu}$ and its covariant form $A_{\mu\nu}$ are given by
\begin{eqnarray}
    &&A^{\mu\nu}=\begin{pmatrix}
        -\frac{4}{9\,\alpha^8\,\varrho^4}\Big(\beta+3\,\alpha^4\,\varrho^2\,(\tau+\beta^2\,z^2)\Big) & 0 & \frac{2}{3\,\alpha^4\,\varrho^2} & -\frac{2\,\beta\,z}{3\,\alpha^4\,\varrho^2}\\
        0 & -\frac{\varrho^2}{3} & 0 & 0\\
        \frac{2}{3\,\alpha^4\,\varrho^2} & 0 & 0 & 0\\
        -\frac{2\,\beta\,z}{3\,\alpha^4\,\varrho^2} & 0 & 0 & -\frac{1}{3\,\alpha^4\,\varrho^2}
    \end{pmatrix},\nonumber\\
    &&A_{\mu\nu}=\begin{pmatrix}
        0 & 0 & \frac{\varrho^2}{6} & 0\\
        0 & -\frac{3}{\alpha^4\,\varrho^2} & 0 & 0\\
        \frac{\varrho^2}{6} & 0 & \frac{t\,\varrho^2}{3}-\frac{\beta}{9\,\alpha^4} & -\frac{\beta\,z\,\varrho^2}{3}\\
        0 & 0 & -\frac{\beta\,z\,\varrho^2}{3} & -\frac{\varrho^2}{3}
    \end{pmatrix}.\label{16}
\end{eqnarray}
At last, the anti-curvature scalar is given by
\begin{equation}
    A=g_{\mu\nu}\,A^{\mu\nu}=-\frac{4}{3\,\alpha^2}.\label{17}
\end{equation}

Thereby, substituting the metric tensor $g^{\mu\nu}$ from (\ref{5}), the anit-curvature tensor $A^{\mu\nu}$ from (\ref{16}), and the anti-curvature scalar from (\ref{17}) into the relation (\ref{13}), we obtain the nonzero components of the symmetric tensor $M^{\mu\nu}=M^{\nu\mu}$ (\ref{13}) given by
\begin{eqnarray}
    \nonumber M^{\tau\tau}&=& \frac{4\,\kappa\,\left(27\,\alpha ^4\,\varrho^2\,(\beta ^2\,z^2+\tau)+7 \beta \right)}{27\,\alpha^8\,\varrho ^4} \, ,\\
    \nonumber M^{\tau\phi}&=& -\frac{2\,\kappa }{\alpha ^4\,\varrho ^2}\, ,\\
    \nonumber M^{\tau z}&=& \frac{2\,\beta\,\kappa\,z}{\alpha ^4\,\varrho ^2}\, ,\\
    \nonumber M^{\varrho\varrho}&=&\kappa\,\varrho ^2 \, ,\\
    M^{zz}&=& \frac{\kappa}{\alpha ^4 \varrho ^2}\, .\label{18}
\end{eqnarray}
The nonzero components of the covariant tensor $M_{\mu\nu}$ are
\begin{eqnarray}
    \nonumber M_{\tau\phi}&=& -\frac{1}{2} \kappa\,\varrho ^2\, ,\\
    \nonumber M_{\varrho\varrho}&=& \frac{\kappa}{\alpha ^4\,\varrho ^2}\, ,\\
    \nonumber M_{\phi\phi}&=& \frac{7\,\beta\,\kappa}{27\,\alpha ^4}-\kappa\,\varrho ^2\,\tau\, ,\\
    \nonumber M_{\phi z}&=& \beta\,\kappa\,\varrho ^2\,z\, ,\\
    M_{zz}&=&\kappa\,\varrho ^2\, .\label{19}
\end{eqnarray}
The trace of the tensor $M^{\mu\nu}$ is given by $M=g_{\mu\nu}\,M^{\mu\nu}=\frac{4\,\kappa}{\alpha^2}$.

In order to simplify the modified field equations (\ref{12}), let's define
\begin{equation}
    J^{\mu \nu} = R^{\mu \nu} +(6\,\alpha^2+\Lambda)\,g^{\mu \nu} + M^{\mu \nu}\, .\label{20}
\end{equation}
Then, the non-zero components of the tensor $J^{\mu\nu}$ using the metric tensor (\ref{5}), the Ricci tensor (\ref{6}), and the tensor $M^{\mu\nu}$ (\ref{18}) are given by
\begin{eqnarray}
    \nonumber J^{\tau\tau}&=& \frac{4\,\left[27\,\alpha^4\,\varrho^2\,(\beta^2\,z^2+\tau)\,(3\,\alpha^4+\alpha^2\,\Lambda+\kappa)+(27\,\alpha ^4+7\,\kappa)\,\beta \right]}{27\,\alpha ^8 \,\varrho ^4} \, ,\\
    \nonumber J^{\tau\phi}&=& -\frac{2 \left(3\,\alpha ^4+\alpha ^2 \,\Lambda +\kappa \right)}{\alpha ^4 \,\varrho ^2} \, ,\\
    \nonumber J^{\tau z}&=& \frac{2 \,\beta \,z \left(3 \,\alpha ^4+\alpha ^2 \,\Lambda +\kappa \right)}{\alpha ^4 \,\varrho ^2}\, ,\\
    \nonumber J^{\varrho\varrho}&=& \varrho ^2 \left(3 \,\alpha ^4+\alpha ^2 \,\Lambda +\kappa \right) \, ,\\
    J^{zz}&=&\frac{3 \alpha ^4+\alpha ^2 \,\Lambda +\kappa }{\alpha ^4 \,\varrho ^2}\, .\label{21}
\end{eqnarray}

Taking $J^{\mu \nu}$ into account, a possible solution will be checked for $\Lambda$ with $\mathcal{T}^{\mu \nu} = 0$ and $\mathcal{T}^{\mu \nu} = \rho\, k ^{\mu}\, k ^{\nu}$, {\it i. e.}, a vacuum and pure radiation field, respectively. One can see from the above set of equation (\ref{21}) that for vacuum case, where $\mathcal{T}^{\mu \nu} = 0=J^{\mu\nu}$,  there is a solution for $\Lambda$ provided the parameter $\beta$ in the space-time (\ref{4}) must be zero, that is, $\beta=0$. Otherwise there is no such solution for $\Lambda$ if the parameter $\beta>0$.

We now focus on the case, where $\mathcal{T}^{\mu \nu} \neq 0$, non-vacuum solution of the modified theory of gravity. In general relativity theory, we have shown that the space-time (\ref{4}) satisfies the field equation with a pure radiation field as matter content with a negative cosmological constant. In the Ricci-inverse gravity, we choose the same pure radiation field as matter content whose energy-momentum tensor is given in equation (\ref{8a}), that is, $\mathcal{T}^{\mu \nu} = \rho \,k^{\mu} \,k^{\nu}$, where in the chart $\{t, \varrho, \phi, z\}$, we defined the null vector field $k_{\mu} = (0, 0, 1, 0)$ and its contravariant form will be $k^{\mu} = (-\frac{2}{a^2\,\varrho^2}, 0, 0, 0)$, such that the vector field satisfies the null condition $k^{\mu}\,k_{\mu}=0$. 

Now, using the relation $J^{\mu\nu}=\mathcal{T}^{\mu \nu}=\rho \,k^{\mu} \,k^{\nu}$ and substituting the non-zero components (\ref{21}) leads to the following system of equation
\begin{eqnarray}
    \nonumber \rho  \left(-\frac{2}{\alpha ^2 \varrho^2}\right)^2\,&=& \frac{4\,\left[27\,\alpha^4\,\varrho^2\,(\beta^2\,z^2+\tau)\{\kappa+\alpha^2\,(3\,\alpha^2+\Lambda)\}+(27 \alpha ^4+7\,\kappa)\,\beta \right]}{27 \alpha ^8 \varrho ^4}  ,\\
    \nonumber 0&=&-\frac{2 \left(3 \alpha ^4+\alpha ^2 \Lambda +\kappa \right)}{\alpha ^4 \varrho ^2}\, ,\\
    \nonumber 0&=&\frac{2 \beta z \left(3 \alpha ^4+\alpha ^2 \Lambda +\kappa \right)}{\alpha ^4 \varrho ^2}\, ,\\
    \nonumber 0&=&\varrho ^2 \left(3 \alpha ^4+\alpha ^2 \Lambda +\kappa \right)\, ,\\
     0&=&\frac{3 \alpha ^4+\alpha ^2 \Lambda +\kappa}{\alpha ^4 \varrho ^2} .\label{23}
\end{eqnarray}
The solution to the above system of equations gives us the following physical quantities
\begin{eqnarray}
    &&\Lambda \to \Lambda_{m} =\Big(-3\,\alpha^2-\frac{\kappa }{\alpha ^2}\Big),\nonumber\\
    &&\rho \to \rho_{m}=\beta\,\left(1+\frac{7 \,\kappa}{27 \,\alpha ^4}\right)\, .\label{24}
\end{eqnarray}

Hence, the space-time described by the line-element (\ref{4}) serves as a solution in modified gravity theories, specifically in Ricci-inverse gravity. In this framework, the matter-energy content consists of a pure radiation field, characterized by a modified energy density $\rho_{m}$ and a modified cosmological constant $\Lambda_{m}$, as given by equation (\ref{24}). These modifications are determined by the coupling constant $\kappa$ along with other parameters ${\alpha, \beta}$.

It is noteworthy that this new theory also allows for the existence of closed time-like curves, which are observed within certain regions where $\tau > 0$. It is worth noting that the modified energy-density $\rho_{m}$ satisfies the energy condition for a positive coupling constant. Notably, the energy condition is automatically satisfied for a positive coupling constant, $\kappa>0$ since $\beta>0$. It can be readily demonstrated that when the coupling constant approaches zero, {\it i. e.,} $\kappa \to 0$, the results obtained revert to the original findings in general relativity, as discussed earlier as well as in Ref. \cite{refa10}.

\section{Conclusions}

General relativity is a classical theory of gravity that has been intensively tested since it was proposed by Einstein. Although Einstein's theory has been successfully verified, it has problems in explaining some observational data. Then, based on observational motivations, alternative models of gravity have been proposed.
Modified theories of gravity have been a subject of significant research interest for an extended period. Several alternative theories of gravity have been proposed by researchers over time. These include the $f(R)$ theory, $f(T)$ theory, $f(R, T)$ theory, $f(R, G)$ theory, and $f(G, T)$ theory (For references to these theories, see \cite{AM}). More recently, a novel gravity theory called Ricci-inverse theory has been introduced in \cite{RIG}. The main characteristic of this theory is that the determinant of the Ricci tensor for any space-time geometry must differ from zero.

In this study, we have examined an example of Petrov type-N pure radiation field solution in the backdrop of anti-de Sitter space. This particular solution allows for the formation of closed time-like curves in specific regions, thus, serving as a model for a time machine within the framework of general relativity. Subsequently, we have taken this type-N space-time and explored it within the context of modified theories using Ricci-inverse gravity. Our findings reveal that this type-N metric is also a solution in this novel Ricci-inverse gravity theory, thus, allows the formation of closed time-like curves, similar to the previous theory. We have observed that the energy-density of the pure radiation field and the cosmological constant undergo modifications due to the coupling constant defined in equation (\ref{24}). It's noteworthy that, as long as the parameter $\beta$ remains positive, the energy-density satisfies the null energy condition as long as the coupling constant is positive. It's worth noting that one can explore matter content other than a pure radiation field, including scalar fields, pressureless perfect fluids, or anisotropic fluids, within the framework of this modified theory of gravity.  In addition, Ricci-inverse gravity is a new theory that should be tested in different contexts. Several points have been investigated, for example, there is no Minkowskian limit for the Ricci-inverse gravity, this theory cannot explain the cosmic expansion history starting from the radiation-dominated epoch to the matter-dominated epoch to the dark energy-dominated epoch, the existence of ghosts must be carefully analyzed, among others. Some studies on instability and perturbation should be considered for future investigation. Therefore, the study developed here follows this necessary line of testing a theory that is proposed as an alternative theory of gravity. It is important to emphasize that general relativity allows solutions that present CTCs that lead to violation of causality. Then verifying an exact solution of general relativity with such a characteristic in theories of modified gravity is an important test like the others mentioned previously. 

The Einstein tensor of a null dust solution (or null fluid) is expressed as $G^{\mu\nu}=\Phi\,k^{\mu}\,k^{\nu}$ \cite{HS,JB}, where $k^{\mu}$ is a null vector field, and $\Phi$ is a scalar multiplier. Introducing a stress-energy tensor in the space-time as $T^{\mu\nu}=\Phi\,k^{\mu}\,k^{\nu} $ satisfies Einstein's field equation, providing a clear physical interpretation in terms of massless radiation. Physically, a null dust solution can describe gravitational radiation or non-gravitational radiation governed by a relativistic classical field theory, such as electromagnetic radiation. Phenomena modeled by null dust solutions include (i) a beam of neutrinos, assumed to be massless and treated using classical physics, (ii) a high-frequency electromagnetic wave, and (iii) a beam of incoherent electromagnetic radiation. Petrov-type N regions are associated with transverse gravitational radiation, the type detected by astronomers using LIGO and Virgo detectors recently \cite{GW1,GW2}. Typically, it experiences decay on the order of $\mathcal{O}(r^{-1})$, indicating that the long-range radiation field falls under type N.

\section*{Data Availability}

No data generated or analyzed in this study.

\section*{Conflict of Interests}

Author(s) declares no such conflict of interests.

\section*{Acknowledgments}

We sincerely acknowledged the anonymous referee's for their valuable remarks and suggestions. F.A. acknowledged the Inter University Centre for Astronomy and Astrophysics (IUCAA), Pune, India for granting visiting associateship. This work by A. F. S. is partially supported by National Council for Scientific and Technological Development - CNPq project No. 313400/2020-2. J. C. R. S. thanks CAPES for financial support.

\global\long\def\link#1#2{\href{http://eudml.org/#1}{#2}}
 \global\long\def\doi#1#2{\href{http://dx.doi.org/#1}{#2}}
 \global\long\def\arXiv#1#2{\href{http://arxiv.org/abs/#1}{arXiv:#1 [#2]}}
 \global\long\def\arXivOld#1{\href{http://arxiv.org/abs/#1}{arXiv:#1}}

\end{document}